\documentclass[authoryear,12pt]{article}

\usepackage{amssymb,multirow,amsmath,graphicx,array,lscape,rotating}
\usepackage{geometry}
\usepackage{amssymb}

\usepackage{natbib}
\usepackage{comment}
\usepackage{color}
\definecolor{dgreen}{RGB}{50,200,0}

\usepackage{array}
\usepackage{colortbl}

\usepackage{makecell}

\begin{document}

\textbf{The value of point of sales information in upstream supply chain forecasting: an empirical investigation}\\
Mahdi Abolghasemi$^1$, Bahman Rostami-Tabar$^2$, Aris Syntetos$^2$\\
$^1$ Department of Data Science \& AI, Monash University, Melbourne, Australia.\\ 
$^2$ Cardiff Business School, Cardiff University, Cardiff, UK.

\begin{abstract}
			Traditionally, manufacturers use past orders (received from some downstream supply chain level) to forecast future ones, before turning such forecasts into appropriate inventory and production optimization decisions. With recent advances in information sharing technologies, upstream supply chain (SC) companies may have access to downstream point of sales (POS) data. Such data can be used as an alternative source of information for forecasting. There are a few studies that investigate the benefits of using orders versus POS data in upstream SC forecasting; the results are mixed and empirical evidence is lacking, particularly in the context of multi-echelon SCs and in the presence of promotions. We investigate an actual three-echelon SC with 684 series where the manufacturer aims to forecast orders received from distribution centers (DCs) using either aggregated POS data at DC level or historical orders received from the DCs.  Our results show that the order-based methods outperform the POS-based ones by 6\%-15\%. We find that low values of mean, variance, non-linearity and entropy of POS data, and promotion presence  negatively impact the performance of the POS-based forecasts. Such findings are useful for determining the appropriate source of data and the impact of series characteristics for order forecasting in SCs.
			
\end{abstract}

Information sharing, Supply chain forecasting, POS data, Promotions, Time series characteristics.
	
\color{black}\section{Introduction}
	
	In the recent decades, different collaborative strategies have been developed to reach potential synergies across the supply chain (SC). There are various studies that discuss different types of SCs, from those exhibiting no collaboration at all, to those associated with collaborative arrangements to enhance forecasting and inventory performance. \citep{singh2018supply,chan2009effect,lau2004impact}. Such collaborative strategies include, but are not limited to, quick response (QR) \citep{perry2001effective}, vendor managed inventory (VMI) \citep{choudhary2015value}, collaborative planning, forecasting, and replenishment (CPFR) \citep{aviv2001effect}, and point-of-sales (POS) information sharing \citep{williams2011top,narayanandemand}.

   	In this paper, we focus on the value of POS information sharing to improve forecasting performance in SCs.  Traditionally, historical orders have been the main source of information for upstream forecasting. More recently, technological advancements have enabled companies to record and share POS data as an expedient source of additional information in SCs. Sharing POS data, in particular, has been put forward as an important mechanism to increase downstream visibility and may be utilized for various purposes such as demand forecasting. Downstream information should be shared with upstream members as much as possible. This can increase visibility across a SC and reduce the associated uncertainty which ultimately can result in more  accurate forecasts \citep{christopher2011supply}.

    POS information can be used for forecasting across different levels of the SCs. Retailers can use them directly to forecast customers' demand. Sharing POS information might be valuable to forecast retailer's demand in downstream SCs \citep{narayanandemand,hartzel2017factors}. 
	POS data, if shared with the upstream SCs, can be used by manufacturers to forecast DCs' orders. 

    There are a number of studies that compare the forecasting accuracy of the POS- and order-based methods in forecasting retailers' and DCs' orders \citep{hartzel2017factors}. While some studies advocate the use of order-based forecasts for DCs' orders \citep{narayanandemand,williams2010creating}, others have found the POS-based forecasts to be more insightful for decision making \citep{croson2003impact}. The literature suggests that various factors such as demand pattern, product type, and inventory policy, may impact the value of POS and order data for forecasting. These factors can be classified into two main categories: i) operational factors, and ii) time series factors.
	
	The first category, operational factors, refers to the products and operational activities of the SC members. This includes product characteristics such as product types and product family, and operational activities and constraints of SC stakeholders such as inventory policy and ordering system \citep{williams2010creating}. 
	The second category refers to the characteristics of the POS and order time series, e.g., demands and orders distribution, trend, seasonality, volumes, promotion status, presence or absence of bullwhip effect, and volatility \citep{croson2003impact,williams2014predicting,hartzel2017factors,narayanandemand}.

   These studies have some limitations which are summarised as follows: i) the impact of time series characteristics on the relative value of POS and order series have not been investigated thoroughly in the presence of promotions ii) the canonical features of POS data and orders including trend, seasonality strength, non-linearity, entropy, hurst (self-similarity or long dependency), stability, skewness, and kurtosis and their impact on forecasting accuracy have been largely ignored iii) the focus of research studies has been on investigating the value of POS data at downstream levels of SCs, i.e., looking at whether using POS data can help retailers to forecast their demand more accurately and decrease their inventory level. However, POS data can be aggregated and shared with the upstream SC members for various purposes such as operations management and strategic planning. The value of POS data in upstream SC has not been investigated enough. iv) there is no rigorous empirical study with a sufficiently large dataset to determine the impact of POS and order series characteristics on forecasting performance. Therefore, there is a need for further empirical studies with large datasets to better understand the value of POS data for upstream forecasting to allow managers to determine whether such information should be used and if so under what circumstances. Our study is motivated by a real world problem and real data collected from 684 stock keeping unit locations (SKULs), making it the largest case study among others.
   
   We address the above discussed gaps by considering a major food company producing fast-moving consumer goods (FMCG) in Australia. Generating accurate demand and order forecasts in the food FMCG helps to minimise stockouts, reduce excessive inventory, and improve availability to customers which all positively impact competitiveness and profitability in organisations \citep{lau2004impact}.
   In order to forecast the manufacturer's demand, we use four different forecasting methods -Autoregressive Integrated Moving Average (ARIMA), Exponential Smoothing in state space form (ETS), Regression with ARIMA errors (R-ARIMA), and ARIMA with explanatory variable (ARIMAX)- and feed them with i) the shared POS data that is aggregated at DCs level, and ii) historical orders of DCs, individually. We use ARIMAX and R-ARIMA methods as our main forecasting models as they take into account external information to generate forecasts and have shown promising results in different forecasting problems \citep{abolghasemi2020demand}. The ARIMA and ETS forecasting methods are popular methods in academia and practice and serve as basic benchmark methods. We also benchmark the predicted demand against the company's forecasts. This is the natural benchmark recognizing though that the main intention here is to derive insights into the comparative performance between POS- and order-based forecasts, rather than making sure that we do better than the benchmarks. We then analyze ten canonical characteristics of the orders and POS series that can describe the series behavior including mean, variance, trend, seasonality strength, non-linearity, hurst, stability, entropy, skewness, and kurtosis to determine their impact on the comparative performance of POS- and order-based series and specify the conditions under which the POS or order series may be more suitable for forecasting purposes. We should note that while it is possible to combine several sources of information such as POS and orders to generate the forecasts, in this study we investigate which source of information can improve the forecast accuracy rather than finding a possible method to combine POS and order information. Combining POS and order information has been studied for pharmaceutical industry \citep{van2020using} but it requires a different study in the retailing context as combining POS and order information in the presence of promotions is a challenging task.
	
	We summarize our contributions to the literature as follows:
	\begin{enumerate}
		\item We investigate the value of POS information to forecast DCs' orders in a SC in the presence of promotions by means of an empirical study in the FMCG industry. 
		
		\item We use aggregated POS data in upstream SC as opposed to downstream SC where POS data is directly accessible.
		
		\item We examine features of POS and order time series including mean, variance, trend, seasonality strength, non-linearity, hurst, stability, entropy, skewness, and kurtosis and develop a hierarchical linear model (HLM) to determine which characteristics have a statistically significant effect on the relative performance of the POS-based to order-based forecasts.
	\end{enumerate}
	
	The remainder of the paper is organized as follows. Section \ref{ISliterature} reviews the literature. Section \ref{case} introduces the case study and describes the dataset of our empirical analysis. Section \ref{methods} explains the experiment setup and Section \ref{ISresults} provides a thorough discussion of the obtained results. Finally, conclusions are drawn in Section \ref{ISconclusion}.

	\section{Information sharing in supply chain} \label{ISliterature}
	
	There are studies that show information sharing, in general, can benefit the SCs by reducing the bullwhip effect \citep{disney2003vendor,wang2016bullwhip}, increasing service levels \citep{claassen2008performance}, saving inventory \citep{lau2004impact}, and reducing cost \citep{lau2004impact,claassen2008performance}. Despite the benefits of information sharing in the SCs, according to two surveys conducted by Capgemini and Forrester Research, only 40\% and 27\% of retailers shared data with their suppliers, respectively \citep{seifert2003collaborative}. Various reasons are reported as obstacles for information sharing such as organizational trust, benefits and cost of information sharing, commitment, and cultural issues \citep{mishra2007information}. In the particular case of sharing POS information, the inability to effectively use the POS data has been a major obstacle for sharing POS information \citep{barratt2003positioning,alftan2015centralised,bassamboo2017inventory}.

	Studies suggest that both operational factors and demand characteristics, e.g., product type, demands and order size, frequency of orders, the variance of orders, and the level of required forecasts in SCs (downstream or upstream), impact the value of POS information for forecasting in SCs \citep{hartzel2017factors,narayanandemand}. 
	The impact of operational factors on the forecasting accuracy of POS- and order-based approaches has been investigated extensively. For example, \cite{hosoda2008there} found that using POS data in a two-echelon SCs, which has order-up-to ordering policy and ARIMA demand, decreases the standard deviation of manufacturer’s demand forecasting error by between 8\% and 19\%, and, consequently, reduces inventory-related costs.
	
    Product types and their demand are commonly investigated as a factor that can impact the value of POS data in forecasting. \cite{williams2010creating} investigated 24 products from three different categories (cereal, yogurt, canned soup) in 18 regional DCs in the USA. They forecasted DCs' orders with naive, moving average, exponential smoothing, and Holt-Winters methods and found that the POS-based forecasts significantly outperformed the order-based ones for cereal products. They also found that the POS-based forecasts is more accurate than the order-based forecasts in 65\% of the observations. Their performance varied across different products with different characteristics. The POS-based forecasts tend to be less accurate for products with large values of bullwhip effect (large ratio of orders variance to POS variance) and high non-turn volume (large promotional sales).
	
	 In one of the few studies that examine the value of POS data for forecasting in upstream SCs, \cite{narayanandemand} evaluated the performance of POS and order data to forecast orders at suppliers and DCs levels, and demand at all levels. They analyzed 21 products (from cereals, detergent, frozen pizza, and cosmetics) and found that the POS-based forecasts do not improve the accuracy of the order forecasts at DCs' level.

		Time series characteristics is another family of factors that can play an important role in determining the forecasting accuracy of methods. Several studies have shown that time series with different features impact the performance of forecasting methods \citep{kang2017visualising,abolghasemi2020demand,spiliotis2020forecasting}. While there are many characteristics that can be extracted from time series \citep{fulcher2017hctsa}, a smaller number of important characteristics can be used to adequately represent the time series behavior \citep{kang2017visualising}. In a comprehensive study on time series characteristics, \cite{wang2009rule} analyzed more than 300 datasets and extracted nine important characteristics of time series including trend, seasonality, periodicity (to determine the seasonality and cyclic patterns), serial correlation, skewness, kurtosis, nonlinearity, hurst (hurst or self-similarity is an index that indicates the long-range dependency in time series), and chaos (chaos or entropy is indicative of time series forecastability which is defined by the level of signals in time series). They applied ARIMA, exponential smoothing, random walk, and neural network methods on their data and exploited the results in the form of six judgmental rules by which the most accurate forecasting method according to the time series characteristics were identified. For example, when time series have a strong trend, high non-linearity, hurst, low seasonality, low kurtosis and skewness, low chaos, and no noise, then the neural network is the most accurate method, followed by ARIMA, exponential smoothing, and random walk. \cite{spiliotis2020forecasting} also used a small set of canonical characteristics to compare the M3 and M4 forecasting competitions datasets. They extracted the entropy, trend, seasonality, linearity, and stability as the primary characteristics of the M3 and M4 series and further exploited skewness, kurtosis, non-linearity, and hurst to describe and compare the behavior of their time series.

	    \cite{hartzel2017factors} examined over 60,000 orders for more than one hundred different items across 25 DCs, and found that using the real-time POS data increases the forecasting accuracy of orders by 11.2\%. They showed that using the POS data improves the accuracy of order forecasts under certain conditions: i) When orders have a low level of frequency, ii) When orders are relatively small or large, iii) When orders have moderate variability. They measured variability by the log of variability of item orders and ranged it on the scale of 0 to 100 percentiles to represent the strength of variability. They indicated the items between 40 to 60 percentiles with moderate variability. They did not provide any specific information about demand characteristics such as trend, seasonality, and promotion status.

	 	We posit that the performance of POS- and order-based forecasts shall be analyzed according to the characteristics of the input series rather than merely their nature, i.e., being POS or order series. We argue that the fundamental reason that makes POS or order series more suitable for forecasting in SCs is not their nature (being POS or order), rather it is their characteristics that render them more suitable for a specific forecasting purpose.
	 	
		When it comes to the value of sharing POS data in SCs, previous studies reveal some limitations that are summarised as follows:
	
	\begin{itemize}

	\item The focus of the studies has been on the downstream SCs, mainly discussing the value of sharing POS data in forecasting retailers' orders. Few empirical studies have been conducted to investigate the value of shared POS data in upstream SCs \citep{narayanandemand,williams2010creating,williams2011top}. 
		
	\item The information-sharing literature has minimally considered the impact of POS and order series characteristics. While the impact of factors such as demand distribution, demand pattern, and products features have been studied, there is no rigorous empirical analysis to investigate the canonical characteristics of orders and POS time series that carry useful information about their behavior. This includes mean, variance, trend, seasonality strength, non-linearity, hurst, stability, entropy, kurtosis, and skewness. We also consider promotion as another important factor. We then evaluate the association between these characteristics and their impact on the accuracy of forecasts.
		
	\item There are mixed results about the benefits of using POS information for forecasting in SCs. While some studies found using POS information can increase the forecasting accuracy of the SCs orders at upstream levels \citep{williams2010creating}, other studies found that using POS is useful for demand planning at retailers' and suppliers' levels \citep{narayanandemand}. More studies are required to investigate the impact of various factors on the forecasting accuracy of the POS-based methods and determine the conditions under which sharing POS data is beneficial for SCs forecasting \citep{williams2014predicting}.
	\end{itemize}

	\section{Case and empirical data}\label{case}
	
\begin{figure}
		\centering
		\includegraphics[scale=0.6] {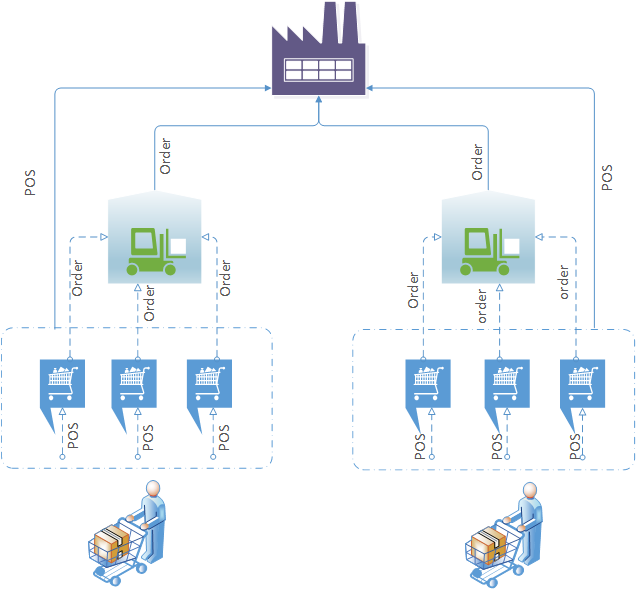}
		\caption{Supply chain and the flow of information}
		\label{scm}
	\end{figure}

	We consider a three-echelon SC of FMCG in Australia with one manufacturer, twelve DCs, and more than 2000 retailer shops to investigate the value of POS information in upstream SC forecasting. While we gather POS information from retailers, our focus is on the upstream SC where we aim to forecast the DCs' orders to the manufacturer. Figure \ref{scm} symbolically illustrates the information flow for our case study SC with one manufacturer, two DCs, and six retailers.

    In our case study SC the retailers place orders to their corresponding DC and DCs place orders to the manufacturer. We look at this problem from the manufacturer's point of view. The manufacturer aims to forecast the DCs’ orders (manufacturer's demand) for production planning, logistics, and inventory management purposes. The manufacturer does not have access to either DCs' operational information such as inventory levels and ordering policy or retailers' orders. Instead, the manufacturer has access to the aggregated-POS data at DCs level that is collected by a third party organization. \textcolor{black}{The aggregated POS data corresponds to 12 DCs each of which includes data for several hundred retailer shops. There are in total 2000 retailers but we do not have the exact number of retailers that correspond to each DC. However, the scale of sales in each DC indicates that some of them have a large market share than others. This data becomes available within a week}. Therefore, the manufacturer has access to two sources of information to generate forecasts: i) historical orders of the DCs, and ii) the aggregated POS data at the DCs level. The manufacturer uses the former in a statistical model (Holt-Winters exponential smoothing) and utilizes the latter source of information to judgmentally adjust the output of the statistical model. Holt-Winters model is a relatively easy model to use and highly interpretable, making it a popular model for industries and our case study to use in practice \cite{syntetos2016supply}. The forecast adjustment is performed by \textcolor{black}{ a team of experts who gather POS and market information from the retailers and use their experience to judgmentally adjust the output of the statistical forecast.The size of the adjustment is larger for promotional periods as the statistical model is not able to effectively incorporate promotional information. This is a cumbersome and cognitively demanding task as experts need to reconcile information and manually produce a large number of forecasts. This is particularly problematic when a promotion is offered as it requires experts to process a large amount of information which may introduce some bias to the forecasts.} In addition, it is not evident which source of information is more suitable for forecasting. In order to automate the forecasting procedure, we use the POS information in a quantitative model that is inspired by the case study problem and tailored to the current practice. We empirically investigate the comparative benefits of using POS and order series for forecasting purposes to help forecasters choose the right source of information and model for forecasting.
	
	We conduct our analysis using data collected from the case study. Our data spans over 112 weeks between September 2016 to October 2018 and includes the weekly aggregated POS series, historical orders, promotion time, promotion type, and price for 684 SKULs. The SKUL refers to a stock keeping unit at different DCs. The SKULs are from different categories and families including breakfast cereals, spreads, and long-life dairy products but their label is unknown for us. These SKULs depict different levels of sales, orders, trends, seasonality, promotion frequency, and promotion type representing a diverse set of POS and order for analysis. 
 
	\subsection{The characteristics of POS and order series}\label{influentialvar}
	
 In order to understand how POS series are different from orders, we investigate their characteristics. While there are many characteristics that can be analyzed for each time series \citep{fulcher2017hctsa}, we exploit ten important characteristics of time series that can explain the behavior of the order and POS series. Identifying and analyzing these characteristics help managers to appreciate how consumers' demands evolve across SCs. These characteristics are defined as follows:

    For more information about calculating these characteristics refer to \citep{wang2006characteristic,kang2017visualising}. We use the `\textit{tsfeatures}' package in R \citep{tsfeatureR} to compute these characterises for POS and order series. 
    
\begin{figure}
		\centering
		\includegraphics[scale=0.6] {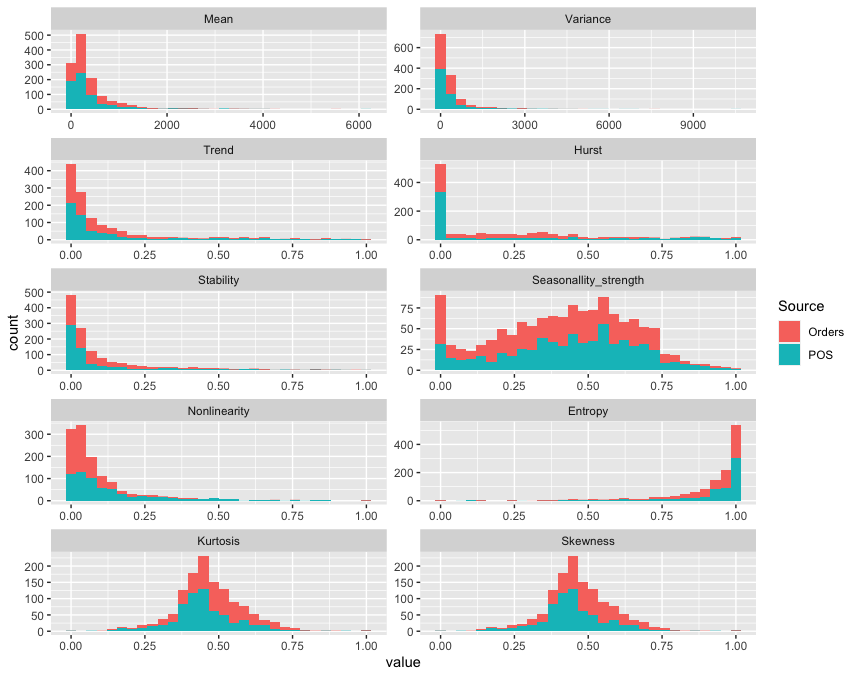}
		\caption{Histogram of order and POS mean, variance}
		\label{kernel}
	\end{figure}

   In order to gain insights on the comparative behavior of order and POS series, we measured the Kullback-Leibler divergence (KL) between order (target probability distribution) and POS series \citep{cover1999elements}. Although KL distance is asymmetric, it can measure how much information is lost when we approximate one distribution with another. Suppose $H(p)$ is the entropy of the target distribution, then we can measure the difference between the estimated and target distribution by $\frac{KL}{H(p)}*100\%$. Table \ref{tsfeatures} provides the descriptive statistics of the order and POS series characteristics for first (Q1), second (Q2), and third (Q3) quartiles and the KL divergence (percentage differences in parenthesis), one overage. These statistics are measured on average across all series. Figure \ref{kernel} shows the histogram of the order and POS series for each characteristic and provides information about the distribution of these characteristics across all series.
	
	\begin{table}[ht]
		\centering
		\caption{Descriptive statistics of order and POS characteristics}
		\begin{tabular}{lccccccr}
			\hline
			{Characteristics}&\multicolumn{3}{|c}{Orders} &\multicolumn{3}{|c}{POS}&\multicolumn{1}{|c}{KL-divergence}\\
			\hline
			& Q1 & Q2 & Q3 & Q1 & Q2 & Q3 & Average\\
			\hline
			Mean  & 140.45 & 271.03 & 511.40 & 92.70 & 198.37  & 408.56& 0.44 (4.23\%)\\ 
            Variance  & 83.55  & 174.91  & 425.77 & 48.86 & 120.69 & 330.55& 0.42 (4.16\%)\\ 
            Trend & 0.00   & 0.03  & 0.08 & 0.01 & 0.04 & 0.15& 0.15 (0.90\%) \\
            Seasonality strength  & 0.14  & 0.26   & 0.36 & 0.21 & 0.32  & 0.42& 0.36 (3.62\%)\\ 
            Non-linearity  & 0.09  & 0.25 & 0.53 & 0.06 & 0.17  & 0.42 & 0.87 (3.98\%)\\ 
            Hurst & 0.15  & 0.19 & 0.28  & 0.15& 0.15  &0.25 & 0.17 (2.69\%)\\ 
            Stability  & 0.01  & 0.05  & 0.15  & 0.01 & 0.03 & 0.17 & 0.06 (0.99\%)\\ 
            Entropy & 0.95 & 0.98  & 1.00 & 0.93 & 0.98  & 1.00& 0.22 (3.19\%)\\
            Skewness & 0.16 & 0.23 & 0.53 & 0.26 & 0.43  & 0.51& 0.33 (2.99\%)\\
            Kurtosis & 0.14  & 0.18 & 0.46 & 0.22 & 0.34  & 0.53  & 0.28 (2.76\%)\\ 
			\hline
		\end{tabular}\label{tsfeatures}
	\end{table}

    In general, both POS and order series depict similar distribution (less than 5\% difference) but different values in Q1, Q2, and Q3. Both POS and order series depict strong entropy. The series do not show strong trend, seasonality strength, non-linearity, hurst, and stability. POS and order series depict relatively small degrees of skewness and kurtosis which are largely caused by promotion.

    The average quantity of order and POS series often vary around each other except during, before, and after promotional periods. Before promotional periods, orders tend to be larger than the POS series, during promotional periods POS is extremely larger than order series, and after promotional periods often POS is larger than orders. The larger amount of orders before promotional periods is due to smoothing orders by DCs and the smaller size of orders after promotional periods indicate that often DCs have enough inventory after promotion and do not place new orders. DCs on average place larger orders to avoid the possibility of stock-out during promotional periods. Note that the smaller values of POS than orders could also be due to the possible stock-out in retailers' inventory but we do not have access to this information for verification.
    
\begin{figure}
    	\centering
    	\includegraphics[scale=0.46] {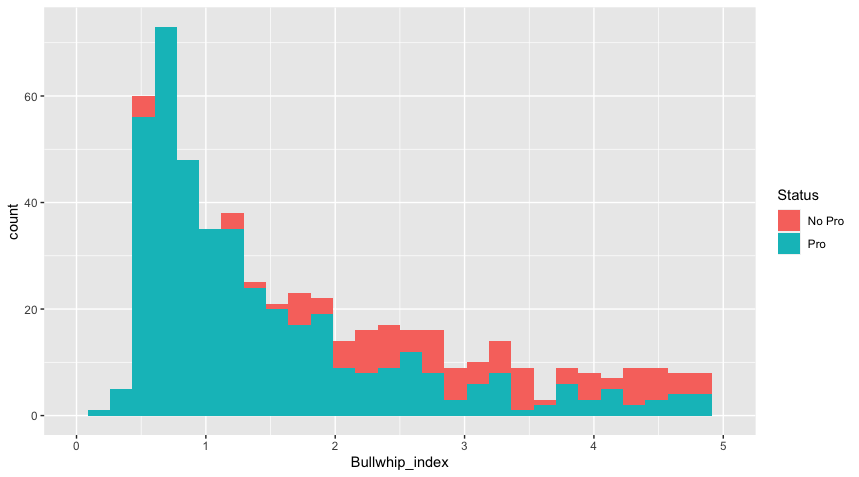}
    	\caption{Bullwhip histogram for different products}
    	\label{bwhisto}
    \end{figure}

    Orders, on average, depict smaller values of the variance than the POS series as it is evident from Q1, Q2, and Q3 of variance for POS and order series. \textcolor{black}{Figure  \ref{bwhisto} shows the histogram of the bullwhip effect for promoted and non-promoted products.} We found that 489 SKULs out of 684 SKULs (71\%) have higher variance in the orders than POS, i.e, they exhibit bullwhip effect phenomenon. More precisely, 196 out of 200 non-promoted series (98\% of the non-promoted SKULs) and 293 out of 484 promoted series (60\% of the promoted SKLUS) exhibit bullwhip  effect phenomenon. There are 195 SKULs (29\% of all SKULs) that showed anti-bullwhip effect behavior, i.e, the variability of orders series was smaller than POS series. The anti-bullwhip effect phenomenon often occurs during promotions when DCs \textcolor{black}{deliberately} smooth their orders during promotions. That is, DCs may place their promotional orders in two periods rather than at once. Therefore, orders show lower levels of variability than the POS series, \textcolor{black}{and consequently the indicative fraction for bullwhip, i.e., $\frac{\sigma^2_{order}}{\sigma^2_{pos}}$, during promotional periods will decrease and contribute to anti-bullwhip behaviour at some scenarios}. This may have impacted their non-linearity and entropy as POS series depict a stronger non-linearity and entropy than the order series. Both order and POS series have shown similar values of hurst. In terms of seasonality, both POS and orders do not depict strong seasonality and POS series have slightly stronger seasonality.  POS and order series have shown similar values of skewness and kurtosis with POS slightly having larger values. The series with large values of kurtosis and skewness are the ones that are heavily impacted by promotions.
    
    In summary, we can see that POS and order series depict different levels of strength in each characteristic. However, our analysis shows that the distribution of the POS and order series are similar at 5\% of significance level in all investigated characteristics. In Section \ref{factors}, we investigate how POS and order characteristics may contribute to the performance of the forecasting models.

	\section{Methods and experiment setup}\label{methods}
	
	In this section, we describe the experimental setup and forecasting methods. We use two types of methods to forecast the DCs' orders. Since some of the order and POS series are impacted by promotion, we use methods with the ability to incorporate promotional information. There are various statistical and machine learning methods in the literature that can be used to model the impact of causal variables. While each method has its pros and cons, there is no unique model that can operate well on all types of data and all sorts of problems \citep{abolghasemi2020demandPromotion}. We choose R-ARIMA and ARIMAX model as two well-known models that have shown promising results in forecasting sales time series, and in particular demand and order time series that are impacted by promotions \citep{abolghasemi2020demand}. Both methods can encapsulate the explanatory variable and incorporate time series dynamic to capture time series variations. We also implement ARIMA and ETS as two basic benchmark models to ensure ARIMAX and R-ARIMA are appropriate forecasting models and outperform simple benchmark models. Both methods use price as the explanatory variable. These methods are discussed in Sections \ref{rarimamethod} and \ref{arimaxmethod}, respectively.

	\subsection{Regression with ARIMA errors}\label{rarimamethod}
	
	R-ARIMA is essentially a regression model that fits an ARIMA($p$,$d$,$q$) model to the residuals to capture any possible undefined correlation in residuals. The parameters $p$, $d$, and $q$ represent the order of Autoregressive (AR) component, differencing, and Moving Average (MA) components, respectively. We use two separate R-ARIMA methods, one uses historical orders of the DCs and the other one uses the aggregated-POS data at the DCs' level. The forecasting model regresses the orders or POS against the price of products and fits an ARIMA process to the errors as shown in Equation \ref{drarima}.
	
	\begin{align} 
	{F_{t}} = \beta_0 + \beta_1 r_{t} + \nu_t,
	\label{drarima}
	\end{align}
	where ${F_{t}}$ denotes forecast at time $t$. The parameter $\nu_t$ is an ARIMA process, and $r_{t}$ is the price at time $t$.
 	
	Note that price carries the promotional information for each product and it is available for the next eight weeks in advance. 
	
	These methods are implemented in R and the `\textit{forecast}' package is used to optimize and fit the parameters of methods \citep{fpp2forecast}.
	
	\subsection{ARIMAX}\label{ARIMAXIS}\label{arimaxmethod}
	We can construct an ARIMAX model by adding an explanatory variable to the well-known ARIMA$(p,d,q)$ model. Price of the products inherently carries promotional information and it is known in advance.  We use price as the explanatory variable in the ARIMAX model shown in Equation \ref{ARIMAXorder}. 
	\begin{align} \label{ARIMAXorder}
	Y_{t} =\beta_1 r_{t} + \phi_1y_{t-1} + ...+ \phi_p y_{t-p} + e_t + \theta_1 e_{t-1} + ... + \theta_q e_{t-q},
	\end{align}
	where $Y_{t}$ is forecast at time $t$, $y_{t-1}$ denotes the actual values of input series at time $t-1$, $r_{t}$ is price at time $t$, and $\beta_1$ is price coefficient. The parameters $\phi$ and $\theta$ are the coefficients of AR and MA components, and the parameters $p$ and $q$ represent the orders of AR and MA components, respectively. 
	
	We use the `\textit{TSA}' package in R to train and estimate the parameters of ARIMAX models \citep{TSA}. Overall, the order of $p$ and $q$ for our fitted models vary between 1 and 4.

	\subsection{Benchmark methods}\label{benchmarks}
	
	We use three benchmark methods in order to illustrate the forecasting performance of the proposed methods. We first compare the forecast accuracy of R-ARIMA and ARIMAX with those of ARIMA and ETS, since these are widely accepted as standard benchmark methods in academia and practice \citep{hyndman2014forecasting}. Both methods are implemented in R and the `\textit{forecast}' package is used to fit and estimate their parameters \citep{fpp2forecast}. \textcolor{black}{We state that ARIMA and ETS models are only used for benchmarking purposes. They may not be appropriate models to capture promotional periods effectively.} Additionally, we benchmark our forecasts against those generated by the manufacturer.

	\subsection{Forecasting accuracy}
	
	We evaluate the forecasting accuracy using the relative mean absolute error (RelMAE) which is a scale-independent accuracy metric \citep{davydenko2013measuring}. Essentially RelMAE is the ratio of the error of the forecasting methods to the baseline method. We calculate RelMAE using Equation \ref{RelMAEformula}.

	\begin{align}\label{RelMAEformula} 
	RelMAE_i = \frac{{\sum_{t=n+1}^{n+h}|Y_{it} -\hat{Y}_t|}}{\sum_{t=n+1}^{n+h}|Y_{bt} -\hat{Y}_t|},
	\end{align}
 	where $h$ is the forecasting horizon, $n$ is the sample size, $Y_t$ is the observed value of the series at time $t$, and $\hat{Y}_{it}$ is the forecast of method $i$ at time $t$. We use ARIMA model (fitted with orders data) as the baseline statistical model.
	
	The first 104 weeks are used as the training set and the last eight weeks are used as the test set to compare the forecasting accuracy of the models. We consider eight weeks ahead forecasts as it is of interest for the company and it provides enough time for manufacturing and logistics plannings. Moreover, price is also available only for the next eight weeks, \textcolor{black}{so we can effectively forecast only for the next eight weeks}. There are 484 of the products that are impacted by promotions and 200 of the products that are not impacted by any promotions. There are 71,608 observations in total from which 9,756 observations (12\% of the periods) are promotional periods. In total, we evaluate 5,472 observation in testset from which 3,923 observations are non-promotional and 1,549 observations are promotional periods. \textcolor{black}{The testset correponds to September and October 2018 and contains data with various features such as trend, seasonality, promotion status, and calendar events; representing a comprehensive dataset for evaluation.}
	We use a pairwise t-test to determine whether there is any significant difference between POS- and order-based forecasts at 5\% of the significance level. 
	
	\subsection{Testing the statistical significance of time series characteristics}
	
	In order to investigate the impact of POS and order series characteristics on the accuracy of POS- and order-based methods, we use a HLM. The HLM suits our data well since demands are nested within each DC. Thus, there might be a relationship between orders of each DC as they may have similar ordering process \citep{hartzel2017factors}. The HLM model replaces the ordinary least squares model which is not suitable in this case since it assumes orders are independent. The HLM takes care of the possible dependency between observations at each DC by considering fixed and random effects on the regression model. The fixed effect examines the direct impact of the explanatory variable and random effect accounts for any possible group effects.  The HLM model is implemented in R and the parameters are estimated using the \textit{lme4} package \citep{lme4}. 
	
	\section{Results and discussion}\label{ISresults}
	
	We present our empirical results in two sections. In Section \ref{foremodels}, we  evaluate the performance of forecasts using either POS or order series and summarise the results. In Section \ref{factors}, we analyze ten characteristics of the POS and order series to identify whether and how they contribute to forecast performance.
	
	\subsection{POS-based vs. order-based forecasts}\label{foremodels}
	
     Table \ref{MASEchp4} shows the obtained results computed based on the mean of RelMAE for each of the discussed forecasting models. The results are summarised across promoted, non-promoted, and all SKULs. We also statistically test whether order-based forecasts have significantly different accuracy from the POS-based ones.

	\begin{table} [!htb]
		\centering	
		\caption{Forecasts accuracy: mean of RelMAE (* significant at 5\% level)}
		\begin{tabular}{lccccccccr}
			\hline
			{Forecasting methods}&\multicolumn{3}{|c}{Promoted} &\multicolumn{3}{|c}{Non Promoted}&\multicolumn{3}{|c}{All SKULs}\\
			\hline
			& Order & POS &  Diff. & Order & POS & Diff. &  Order & POS & Diff. \\
			\hline
			ARIMAX &  0.87 & 0.98  & $11\%^*$ & 0.88 & 1.04 & $15\%^*$ &0.88 &1.03 & $14\%^*$\\
			R-ARIMA &  0.88 & 1.01  & $12\%^*$ & 0.89 & 1.04 & $14\%^*$ & 0.89 & 1.04 & $14\%^*$ \\
			ARIMA &  1.00 & 1.03  & 3\% &  1.00 & 1.10 & $10\%^*$ & 1.00 & 1.04  & 4\%\\
			ETS &  1.01 & 1.04  & 4\% & 1.01  & 1.07 & $6\%^*$ & 1.02 & 1.06  & 3\%\\
			Company's model & 0.96 & - & - & 0.99 &-   & -& 0.99 & -& -\\
			\hline
			\label{MASEchp4}
		\end{tabular}
	\end{table}
 
    Our results show that the order-based forecasts, on average, outperform the POS-based forecasts in forecasting the DCs' orders by between 6 to 15\%. We observe that the superiority of order-based models decreases in the presence of promotions. For the promoted SKULs, the results show that the order-based forecasts significantly generate more accurate forecasts (at 5\% significance level) than the POS-based forecasts when the ARIMAX and R-ARIMA methods are utilized to forecast the DCs' orders. However, when the ARIMA and ETS methods are employed to forecast the DCs' orders for promoted products, there is no significant difference between the accuracy of the forecasts that use either order or POS series. This is because neither of the univariate ARIMA and ETS methods incorporates promotional information to forecast the DCs' orders and these methods may not be suitable to forecast our order and POS series that are impacted by promotion. Consequently, they both have large forecasting errors and their forecasts are not significantly different. For the non-promoted SKULs, there is a statistically significant difference at a 5\% level between the forecast accuracy of all employed methods when using POS or order series.
	
	The ARIMAX and R-ARIMA methods depict higher accuracy than the ARIMA and ETS methods. This is of no surprise since both the ARIMAX and R-ARIMA methods use price as additional information to forecast orders as opposed to ARIMA and ETS that are univariate methods and only use previous values of series. This indicates that additional information about price can be useful only if it is added and modeled properly to the model.
	The accuracy of the company's forecasting method, on average, is superior to the ARIMA and ETS methods. This implies that using POS information in a subjective manner (in the form of expert's judgments) adds value to the company's statistical forecasts. However, the ARIMAX and R-ARIMA methods outperform the company's method, both in the absence and presence of promotions, indicating that choosing an appropriate type of forecasting method that can effectively and objectively take into account the POS information is beneficial.
	
	We also investigate the frequency of the periods that one model outperforms the other one. The order-based forecasts using the ARIMAX and R-ARIMA methods outperform the POS-based ones over 68\% and 61\% of the periods, respectively. This ratio drops to 51\% and 49\% for the ARIMA and ETS methods, respectively. In order to know what factors make one model more accurate than the other one for a particular series, we analyze their behavior using various characteristics as discussed in Section \ref{influentialvar} and provide a thorough discussion in Section \ref{factors}.
	
	Our results with regards to the superiority of order-based forecasts to the POS-based ones in upstream SCs are consistent with recent findings by \cite{narayanandemand}. However, our experiment differs from their study and adds value in the following ways: i) while their finding is only based on 21 products and includes partially synthetic data, we use a large real-world dataset with 684 SKULs, ii) POS and order series used in our study contain different features, depict different behavior and includes promotions as an important feature, and iii) we employ different forecasting methods in our experiment.
	
	We summarise our findings as follows:  First, it is imperative to choose the right forecasting method to be able to compare the value of the POS and order series. We showed that if we choose inappropriate models to forecast the series, there will be no statistically significant difference between the accuracy of the forecasts regardless of the chosen input series. Second, for an appropriate choice of a forecasting model, using order series results in higher forecast accuracy than using the POS data to forecast the DCs' orders. However, this superiority declines in the presence of promotions and may depend on the behavior and characteristics of the series. Third, in our case study we observe that using POS data in a judgmental fashion can improve the forecasting accuracy of a statistical model that is merely based on historical order series. However, POS data can be used more objectively in a quantitative model to forecast the manufacturer's demand. This can result in higher forecasting accuracy without any intervention.

	\subsection{Discussion on the characteristics of order and POS series}\label{factors}
	
	The results of the M4 forecasting competition and other studies verify that the characteristics of time series impact the forecasting accuracy of methods \citep{wang2009rule,kang2017visualising,spiliotis2020forecasting}. 	Now we turn our attention to determine the significant characteristics of the POS and order series and their corresponding impact on the accuracy of the forecasts. It is of great importance to know the characteristics of the POS and order series and circumstances under which using one source of data is superior to one another in forecasting orders. We found that there is no significant difference between the forecasting accuracy of the ARIMAX and R-ARIMA methods as two top-performing methods when the same source of data, i.e., either POS or order series, is used to generate forecasts. Therefore, we only analyze the forecasts generated by the ARIMAX method.

   We evaluate the performance of the POS-based and the order-based forecasts and investigate their association with POS and order characteristics using an HLM model. We measure the relative accuracy of order-based to the POS-based forecasts by $MAE(POS)/MAE(Order)$, and regress this dependent variable against the characteristics of all POS and order series as fixed independent variables. Although we can not calculate these characteristics for the promotional and non-promotional observations individually, we take into account the promotion frequency and promotion impact by considering the ratio of promotional periods as another independent variable.  Note that the promotion ratio for POS and order series is identical and it is indicated with a single variable in the HLM model. We fit one global model on all 684 series characteristics and report the results. Table \ref{DCordpos} shows the variables, their estimated coefficient, standard error, t-value, and their p-values. We obtain the p-values with the likelihood ratio test where we tested the full model against the reduced model of the variable that we are interested in.

	\begin{table}[h!]
		\centering
		\caption{The HLM model results (\**significant at 5\% level, \*** significant at 10\% level)}
		\begin{tabular}{llllll}
			\hline
			& Input &Coefficient & Std. Error &  t value & P-values \\ 
			\hline
			(Intercept) & & 1.13 & 0.29  & 3.89 & $0.00^*$ \\ 
			Mean &POS& -0.21 & 0.09  & -2.31 &  $0.02^*$ \\ 
			& Orders & 1.51 & 0.84 &  1.79 & $0.07^{**}$ \\
			
			Variance & POS & -0.29 &  1.55  & -0.18 &  $0.01^*$ \\
			& Orders &  0.64 & 1.49 &  0.43 & $0.05^*$ \\
			
			Trend & POS& 0.17 & 0.26 &  0.67 & 0.50 \\ 
			& Orders & -0.06 & 0.22 &  -0.30 & 0.76 \\ 
			
			Seasonal strength & POS & 0.13 &  0.07 & 1.69 & 0.09 \\ 
			&Orders &  0.01 &  0.08 & 0.15 & 0.87 \\ 
			
			Non-linearity &POS & -0.38 & 0.20 &-1.90 & $0.06^{**}$ \\ 
			&Orders &  0.20 & 0.09 & 2.06 &  $0.04^*$\\ 
			
			Hurst & POS & -0.18 & 0.11 & -1.57 & 0.11 \\ 
			& Orders &-0.03 & 0.09 &  -0.33 & 0.73 \\ 
			
			Stability & POS & -0.29 & 0.27 & -1.05 & 0.29 \\
			& Orders &0.15 & 0.24 &  0.64 & 0.52\\ 
			
			Entropy &POS & -0.23 &  0.20 & -1.15 & $0.06^{**}$ \\ 
			& Orders & 0.01 &  0.22 & 0.03 & $0.07^{**}$ \\
			
			Skewness &POS& 0.01 & 0.16  & 0.08 & 0.93 \\ 
			& Orders &  0.03 & 0.15 &  0.21 & 0.82 \\

			Kurtosis & POS & 0.29 & 0.12 &  2.46 & 0.14 \\
			& Orders &  -0.22 & 0.11 &  -1.91 & 0.55 \\
			
			Promotion & POS & -0.05 & 0.02 & -2.21 & $0.03^*$ \\ 
			\hline
		\end{tabular}\label{DCordpos}
	\end{table}
	
    Our research findings demonstrate that the mean of POS series, the variance of the order and POS series, the entropy and non-linearity of order and POS series, and the frequency of promotions are the characteristics of the series that have a statistically significant impact on the comparative performance of the POS and order data in forecasting DCs' orders. In our experiment, the other investigated characteristics of series including trend, seasonality strength, hurst, stability, kurtosis, and skewness of POS and order series do not significantly impact the relative performance of the POS- to order-based forecasts. 	
    
	According to the results in Table \ref{DCordpos}, the relative MAE of the POS-based forecasts to order-based forecasts decreases when the POS mean increases, i.e., the relative performance of POS-based forecasts increases as POS mean increases. We attribute this to the intermittency nature of order series when the POS series have a small mean. We scrutinized the POS and order series and found that the SKULs with very low volumes of POS correspond to the SKULs with lower levels of orders frequency. For these series, there is no order placed at some periods, i.e., order series have intermittent nature with low intermittency level. Since the DCs' orders depict intermittency, using POS data to forecast the DCs' orders may not be suitable. This is consistent with the results that are reported at retailers level. According to \citep{williams2010creating}, the higher volume of sales improves the relative performance of the POS-based methods to the order-based methods to forecast retailers' orders.
 	
    The variance of the POS and order series are the other statistically significant characteristics of the series that can impact the relative performance of POS- and order-based forecasts. Our results show that the relative MAE of the POS-based to order-based forecasts is negatively correlated with the variance of the POS series. That is, the relative performance of the POS-based to order-based forecasts increases when the variability of the POS series increases. This may attribute to the promotion impact and the order smoothing during promotions that cause anti-bullwhip behavior in series. When POS series are impacted by large promotions they exhibit large variability. The corresponding order series depict smoother behavior and smaller variance as DCs smooth their orders during promotional periods. However, DCs change their orders frequently to match their actual demand. As such, the orders placed by the DCs may be disturbed and they may not be a good signal for forecasting DCs' orders, instead, POS information can be more helpful to forecast the uncertain orders.  The same results hold for non-linearity, i.e., the relative MAE of POS-based to order-based forecasts decreases when the non-linearity of the POS series increases, i.e., the performance of POS-based forecasts increases as POS variability increases. Non-linearity in POS series is mainly caused by promotion and our results show that when series depict higher values of non-linearity, it is better to use POS series that directly carry promotional information rather than using DCs' orders that are disturbed by DCs' forecasts error and operational constraints.

   The entropy of POS and order series also found to be another characteristic of the POS and order series that  significantly impact their forecasting accuracy. Our results show that the relative MAE of the POS-based to order-based forecasts is negatively correlated with the entropy in the POS series. That is, the POS-based forecasts tend to perform better when the entropy of POS series increases. This may attribute to the DCs' smoothing during promotions and in general DCs' forecasting accuracy. Since noisy data are less forecastable, DCs may have less accurate forecasts when POS data are noisy and accordingly change their orders to match with their actual demand. DCs and retailers are not usually as good as their suppliers in forecasting\citep{gilliland2010business}. As such, the orders placed by the DCs may be disturbed and they may not be a good signal for forecasting. Thus, it might be better to rely on POS series for forecasting.

    According to our results, the negative coefficient of the promotion factor indicates that the relative MAE of the POS-based forecasts to order-based forecasts decreases in the presence of promotions, i.e., the performance of POS-based forecasts increases in the presence of promotion. One explanation is that orders and POS series are more similar in the absence of promotion. However, DCs smooth their orders during promotions and POS data at each period does not directly translate to the orders but span over a few periods with uncertain values that depends on operational factors and ordering policy of the DCs. This may render POS data as an appropriate candidate to forecast the DCs' orders during promotions. This finding aligns with the current practice in the company and experts’ opinions who state that POS information is a valuable source of information to forecast DCs' orders during promotions. This finding may limit to our study and attribute to promotion and the specific ordering policy of DCs that occurs during promotions. The promotions impact on the value of the POS and order series at different stages of the SCs need to be further investigated.

   In summary, we show that using the order series to forecast manufacturer's demand, on average, results in higher forecast accuracy than using the POS series. However, their relative performance is tied to the characteristics of the input series and may change from one series to another. We found that the mean, variance, non-linearity and entropy of POS series, and promotion status are statistically significant features that can impact the value of the POS data for forecasting orders. We state that the usefulness of the POS or order series for forecasting may change across the SCs and managers should not evaluate their value merely based on their nature, i.e., being POS or order. \textcolor{black}{Our results show that POS data may be more adequate for order forecasting in the presence of promotion, otherwise historical orders are more appropriate for forecasting non-promotional orders. Also, one should consider various features of POS data such as mean, variance, non-linearity, and entropy to determine whether they willbe useful for order forecasting or not. Table \ref{summary} shows the summary of results. As shown in Table \ref{summary}, the relative performance of the POS-based to order-based forecasts tend to improve under five conditions: i) the mean of POS series increases while orders are not intermittent, ii) the variance of POS series increases, iii) the entropy of POS series increases, iv)  the non-linearity of POS series increases, v) promotions are offered to customers.  }
   
   	\begin{table}[h!]
   	\centering
   	\caption{Summary of findings with regards to relative performance of POS to order based forecasts}
   	\begin{tabular}{lll}
   		\hline
   		Feature & Status &POS/Order \\ 
   		\hline
   		POS Mean &Increasing & Increasing \\ 
   		POS Variance & Increasing & Increasing \\
   		POS entropy & Increasing & Increasing \\    	 
   	 	POS non-linearity & Increasing & Increasing \\ 
   	 	Promotion & On & Increasing \\
   		\hline
   	\end{tabular}\label{summary}
   \end{table}
   

	\section{Conclusions}\label{ISconclusion}
	
	In this paper, we empirically analyze a three-echelon SC from a major FMCG food company in Australia. We focus on the upstream SC where the manufacturer aims to forecast the DCs' orders. DCs do not share their operational information such as ordering policy and inventory level with the manufacturer but the retailers POS data is available in the aggregated form at DCs level. The manufacturer can use either historical orders or aggregated POS data to forecast the DCs' orders. The order and POS series are highly impacted by promotions and depict different behaviors.

	In order to investigate the value of POS information sharing in upstream SC forecasting, we compare the accuracy of POS- and order-based methods in forecasting the DCs' orders. We empirically show that the order-based forecasts outperform the POS-based forecasts by between 6\%-15\% (on average) in forecasting the DCs' orders. We illustrate that the comparative value of the POS and order data depends on the type of forecasting method and series characteristics. 
	
	We use four different forecasting methods to forecast the DCs' orders. Since POS and order series are impacted by promotion, we use the ARIMAX and R-ARIMA methods to forecast the DCs' orders and compare them with the common benchmark methods, ARIMA and ETS. We empirically analyze their performance using the orders and POS series for 684 SKULs and show that our proposed methods can effectively take into account the POS information to forecast the DCs' orders. We also compare our model with the employed model in the case study and show that our proposed ARIMAX and R-ARIMA methods outperform the company's forecasts that incorporate the POS data subjectively into the output of its order forecasting model.
	
 	We further analyze ten important characteristics of the orders and the POS time series to determine how information evolves across the SCs. We show that the distribution of orders and POS are similar in various characteristics but they depict different levels of strength. We determine the significant characteristics of the POS and order series that contribute to their forecasting performance. We illustrate that order-based models are, on average, more accurate than the POS-based ones in upstream SCs. \textcolor{black}{However, the performance of the POS based forecasts tend to improve during promotions and when POS series depict higher mean, variance, entropy, and non-linearity.}
	 
	The results of this research shed light on the value of the POS data in upstream SCs forecasting and provide guidelines for practitioners on using the suitable types of input series for forecasting DCs' orders. There are some limitations to this study that may be considered for future research. \textcolor{black}{Primarily, our insights are tied to the current dataset that is mostly cereal and breakfast product with an FMCG nature. Although the models and results are analyzed on a large dataset to ensure its robustness and validity for dataset with similar characteristics, the resuts may not be applicable for other datasets such as slow moving products. Thus,  replicating the experiment on other datasets with different characteristics may reveal new insights.} In addition, \textcolor{black}{in this study we focused on weekly dataset but a natural extension to this study is to consider temporally aggregated data, e.g., monthly, and investigate whether the same results hold or not.  We also suggest that }theoretical analysis can be beneficial to build theories in this area. The other important research avenue is to further evaluate the promotions impact. While promotions have been extensively studied in the forecasting literature, it is less considered in the information sharing literature and needs to be further investigated both in theoretical and empirical research. The utility of forecasts in SCs is another important research direction. The higher accuracy of forecasts can impact SCs performance in different ways and eventually result in a lower inventory level, higher service level, and lower cost. However, the improvement in forecasting accuracy does not necessarily reduce the inventory costs and the relationship between these two need to be investigated more \citep{syntetos2006stock}.
	
	Finally, we use either the POS or order series individually to forecast DCs' orders. \cite{williams2014predicting} used both POS and order data to predict retailer's orders and showed that their approach improves the forecasting accuracy by 125 \%. They found that the retail echelon inventory processes translate into a long-run equilibrium between order and POS called the inventory balance effect. However, their approach does not accommodate promotions impact and is not suitable if a cointegration does not exist for POS and order series. The presence of promotion and order smoothing changes the dynamic of POS and order series, resulting in conditions where cointegration of them may not be stationary. \cite*{van2020using} investigated 50 products from the pharmacological industry and found that using both sell-through (POS) and order series can improve the forecast accuracy at the manufacturer's level by 4.7\%, on average, and at most by 10\%. Different results are reported in the literature and more studies are required to examine the value of using both POS and order series simultaneously to forecast DCs' orders.
	
	\newpage
	\bibliographystyle{agsm}
	\bibliography{aggrcomp.bib}
	
	\newpage
	
	\appendix
	
	\section{Time series Characteristics}\label{time_series_characteristics}
	
		\begin{enumerate}
		
		\item Mean: Mean is the arithmetic average of a time series. 
		
		\item Variance: Variance is the spread of observations in a time series. We calculate the variance of orders and POS for each SKUL. The ratio of orders variance \textcolor{black}{($\sigma^2_{order}$)} to POS variance  \textcolor{black}{($\sigma^2_{pos}$)} indicate the bullwhip effect index for a particular SKUL \citep{disney2003vendor}.

		\item Trend: Trend refers to a process where time series mean changes over long term periods. Consider a time series $Y_t=S_t+T_t+E_t$, where $S_t$, $T_t$, and $E_t$ denote the seasonal, trend, and errors, respectively. We can detrend  and deseasonalise a time series by subtracting the trend, and seasonality components from series. Let $X_t=Y_t - T_t$ and $Z_t=X_t - S_t$ represent the detrended and deeseasonalised components, respectively. Then, the remainder is error term, $e=Y_t - T_t - S_t$. Trend strength can be measured by Equation \ref{trend.strength} \citep{pollock2001methodology,wang2009rule}.
		\begin{align}
		\text{Trend strength} & = 1 - \frac{\operatorname{Var}(e_t)}{\operatorname{Var}(Z_t)} \label{trend.strength} \\ \nonumber
		\end{align}
		
		\item Seasonality strength: Time series depict seasonality pattern when they exhibit a pattern that is repetitive and it is often caused by seasonal factors. Time series with a fixed seasonality have a large autocorrelation at fixed seasonal lags. Seasonality strength can be  measured by Equation \ref{seasonal.strength} \citep{wang2009rule}.
		
		\begin{align}
		\text{Seasonality strength} &  = 1 - \frac{\operatorname{Var}(e_t)}{\operatorname{Var}(X_t)}
		\label{seasonal.strength}
		\end{align}

		\item Non-linearity: This metric measures the strength of non-linearity relationship between observations of a time series. We measured non-linearity with the modified version of the Teräsvirta’s non-linearity test as described in \citep{tsfeatureR,terasvirta1993power}.

		\item Hurst: Hurst measures the long term dependence of a time series. 
		This measure can be estimated with different methods \citep{rose1996estimation}. In order to compute hurst, an autoregressive fractionally integrated moving average model is fitted to the time series where $d$ is the degree of first differencing \citep{tsfeatureR}. Hurst is equal to 0.5 plus the maximum likelihood estimate of the fractional differencing order $d$ to make it consistent with hurst coefficients \citep{wang2009rule}.
		
		\item Stability: Stability measures the variance of means of series. We first calculate the mean of POS and order for each series, then calculate the variance of the means across all series. This metric gives statistics about the spread of means of POS and order series.

		\item Entropy: This measures the “forecastability” of a time series. Large values of entropy indicate a high noise-to-signal ratio that makes them more difficult to forecast \citep{goerg2013forecastable}. 
		Entropy is the \textit{Shanon entropy} as shown in the Equation \eqref{entropy},
		\begin{align}
		\label{entropy}
		\text{entropy}= - \int_{-\pi}^{\pi} \hat{f}(\lambda) \log(\hat{f}(\lambda)) d\lambda, 
		\end{align}
		where $\hat{f}(\lambda)$ is the spectral density of the data. The spectral density describes the power or strength of a time series as a function of frequency $\lambda$.
		
		\item Skewness: Essentially skewness measures the extent by which a distribution is deviated from a normal distribution. We measured the skewness using Equation \ref{skewness}:
		\begin{align}
		\label{skewness}
		\text{Skewness}= \frac{1}{n\sigma^{3}}\sum_{t=1}^{n}(Y_t - \bar{Y})^3,
		\end{align}
		where $n$ is the number of observations, and $\bar{Y}$ and $\sigma$ represent the average and standard deviation of the time series, respectively.
		
		\item Kurtosis: Kurtosis measured whether the distribution of data is heavy-tailed or light-tailed. Higher values of kurtosis represent time series with heavy tails and lower levels of kurtosis indicates time series with light tails.
		We measure kurtosis with Equation \ref{kurtosis} as shown below:	
		\begin{align}
		\label{kurtosis}
		\text{Kurtosis}= \frac{1}{n\sigma^{4}}\sum_{t=1}^{n}(Y_t - \bar{Y})^4,
		\end{align}
		where $n$ is the number of observations, and $\bar{Y}$ and $\sigma$ represent the average and standard deviation of the time series, respectively.
	\end{enumerate}

\end{document}